\documentclass[12pt,twoside,dvips]{article}
\usepackage{amssymb}
\usepackage{amsfonts}
\usepackage{amsmath}
\usepackage{pifont}
\usepackage[spanish,english]{babel}
\usepackage{graphicx}

\begin{document}

\title{$Z^{\prime}$ boson detection in the Minimal Quiver Standard Model}
\author{D. Berenstein$^{1}\thanks{%
e-mail: dberens@physics.ucsb.edu}$, R. Mart\'{\i}nez$^{2}\thanks{%
e-mail: remartinezm@unal.edu.co}$, F. Ochoa$^{2}\thanks{%
e-mail: faochoap@unal.edu.co}$ and  S. Pinansky$^{3}\thanks{%
e-mail: samuelp@hep1.c.u-tokyo.ac.jp}$ \and $^{1}$Department of Physics, University of California at Santa Barbara, CA 93106 \and $^{2}$ Departamento de F\'{\i}sica, Universidad
Nacional de Colombia, Bogot\'{a} D.C.\and $^{3}$Institute of Physics, University of Tokyo, \\
Komaba, Meguro-ku, Tokyo 153-8902 Japan 
}
 
\maketitle

\begin{abstract}
We undertake a phenomenological study of the extra neutral Z' boson in the Minimal Quiver Standard Model and discuss limits on the model's parameters from previous precision electroweak experiments, as well as detection prospects at the Large Hadron Collider at CERN. We find that masses lower than around $700$ GeV are excluded by the $Z$-pole data from the CERN-LEP collider, and below $620$ GeV by experimental data from di-electron events at the Fermilab-Tevatron collider. We also find that at a mass of 1 TeV the LHC cross section would show a small peak in the di-lepton and top pair channel.  
\end{abstract}

\section{Introduction}

Many extensions of the Standard Model (SM) predict new massive, neutral gauge
particles called $Z^{\prime }$ bosons (for reviews see \cite{zprimas, Langacker:2008yv}). The detection of a $Z^{\prime}$
resonance has become a matter of high priority to particle physics,
because it could reveal many features about any underlying unified theory.
Indirect searches for these neutral bosons have been carried out at LEP by looking for mixing with the $Z$ boson \cite{Alcaraz:2006mx}. Experiments into directly producing $Z^{\prime}$ bosons
are being performed at the Tevatron \cite{Abulencia:2006iv}-\cite{top}. The potential for discovering $Z^{\prime}$ particles at the forthcoming Large
Hadron Collider (LHC) has been explored in the $M_{Z^{\prime}}\approx 1-5$ TeV range \cite{Dittmar:2003ir}-\cite{hp0707.2066}.
The search for this particle using the planned
International Liner Collider (ILC) has also been explored \cite{ILC}.

There are many theoretical models which predict a $Z^{\prime}$ mass in the
TeV range, where the most popular are the $E_{6}$ motivated models \cite{zprimas,DelAguila:1995fa},
the Left-Right Symetric Model (LRM) \cite{Mohapatra:1986uf}, the $Z^{\prime}$ in Little Higgs
scenario \cite{ArkaniHamed:2001nc}, and the Sequential Standard Model (SSM), which has heavier couplings than those of the SM Z boson. Searching for $%
Z^{\prime}$ in the above models has been widely studied in the literature \cite{zprimas} and
applied at LEP2, Tevatron and LHC (a  recent review on the current status of $Z'$ physics can be found in \cite{Langacker:2008yv}).  In particular, a string inspired model was recently proposed  which predicts a single extra $Z^{\prime}$ boson with specific couplings to the standard model matter \cite{Berenstein:2006pk}.

This Minimal Quiver Standard Model (MQSM) is the simplest quiver gauge theory that both contains the standard model and which could arise as the low-energy effective gauge theory of a perturbative D-brane model in string theory.  Stacks of N coincidental D-branes can give rise to gauge groups of the form $U(N)$, $SO(N)$ and $Sp(N)$ (those with a large $N$ limit), and strings stretching between these branes carry charges in the bifundamental representations of these groups (a review of string constructions can be found in \cite{BCLS}).  Quiver diagrams give a pictorial representation of these setups. The quiver diagram assigns a graph with marked edges to a given field theory. The nodes represent gauge groups and the edges represent matter with fundamental or antifundamental representations of the gauge group at each end of the edge (these are some of the low lying open string states connecting the stacks of branes). An edge with two ends of the same group is a tensor with two indices (each of which can be fundamental or antifundamental) under the gauge group.
The effective low energy theory of this effective brane-world scenario is a quiver gauge theory with only bifundamental matter fields.  Simply enumerating the smallest possible quivers shows that the simplest example of a quiver gauge theory that could come from a D-brane model is the one shown in figure 1 with gauge group $U(3)\times Sp(1)\times U(1)$, which we will call the MQSM.  In the figure, each node represents a gauge group and each arrow a fermion transforming in the bi-fundamental representation of the nodes to which it is attached.  The Higgs can also be accommodated as an extra scalar field, shown as a dashed line in the figure.
The labels of the edges indicate the corresponding matter field names in the ordinary standard model ($q_L$ indicate the quark doublets, $\ell_L$ indicate the lepton doublets, etc). The $SP$ and $SO$ groups arise from systems with unoriented strings. These are also usually called orientifolds.   This enables us to obtain the right hypercharges for both right and left-handed quarks with only 3 total gauge groups, because the string modes for up and down quarks might have different orientations at the non QCD end.  A summary of one generation of the particle spectrum is given in table 1.

A careful analysis of the anomalies of the theory shows that one of the two overall $U(1)$'s is the standard hypercharge, and the orthogonal combination is anomalous.  In a pure gauge theory, this would be a serious problem. However in the context of an effective D-brane model this mixed anomaly is cancelled by a Green-Schwarz mechanism. This requires an extra axionic particle which arises from the closed string gravity sector and which transforms inhomogeneously under the gauge transformations.  This has the side-effect of giving an explicit mass term for the associated gauge boson which is dependent on the specifics of the background geometry of the string theory.  Hence the MQSM has one single particle beyond the Standard Model, a massive neutral gauge boson. These extra $Z'$ bosons can not be avoided in D-brane models \cite{GIIQ} (see also \cite{CIK}). Moreover, the anomaly cancellation mechanism 
requires dimension five operators in the Lagrangian, so a direct observation of such a $Z'$ particle would predict a low string scale in order to unitarize the theory at higher energies.

This model has a number of phenomenological advantages which makes it interesting to study.  First, it is extremely simple, with only a single extra particle beyond those in the SM, which enables calculations to be done exactly.  Second, it has only two unknown parameters, the mass of the $Z^\prime$ and the mixing angle between the two $Z$'s (which are related to the Green Schwarz mass term and string energy scale in the underlying theory).  Third, unification of the extra $U(1)$ factor with the color $SU(3)$ causes the $Z'$ to be leptophobic, which allows it to remain phenomenologically viable at TeV scale masses even after electroweak constraints are taken into account. We will explain this fact later on. 

In this work we report a phenomenological study of the MQSM neutral boson. First, we consider indirect limits at the $Z$ resonance obtaining allowed regions for the $Z-Z^{\prime}$ mixing angle and the $Z^{\prime}$ mass. The above analysis is performed through a $\chi ^{2}$ statistics at $95\%$ C.L., including correlation data among the observables. Later, we search for $Z^{\prime}$ bosons
in di-electron events produced in $p\bar{p}$ and $pp$ collisions at Fermilab-Tevatron
and CERN-LHC colliders, respectively. We also search for event signals in the top channel at LHC.

\section{The Model}
The standard model group $SU(3)\times SU(2)\times U(1)\sim U(3)\times Sp(1)$ can be accommodated in a two-node quiver model that is only a product of two groups . However, the matter content can not, as the leptons are doublets under $Sp(1)$ and are not charged under color.  This means that in D-brane models we need to extend the gauge group of the standard model. The minimal extension requires us to enlarge the gauge group by the smallest amount possible,
 giving a gauge group $U(3)\times Sp(1)\times U(1)$, i.e. the model has three stacks of branes. The other possibility of $U(3)\times U(2)$ cannot accommodate right handed quarks with different hypercharge. The insight that the $SU(2)$ weak group can be described as $Sp(1)$ in string models has been advocated in \cite{Cremades:2003qj}, as this reduces the required number of Higgs doublets to generate all Yukawa couplings at tree level.
 
Using this gauge group there is only one choice for the left handed quark doublet $q_L$, a bifundamental $(3,2)_0$.  Since we know the right handed quarks have different hypercharge we make $\overline{u}_R$ transform as $({\bar 3},1)_1$ and $\overline{d}_R$ as $({\bar 3},1)_{-1}$. As described in \cite{Beren}, if we make the right handed quarks appear as two index representations of $SU(3)$ (like they do in $SU(5)$ GUT models), the Yukawa couplings for some of the quarks are forbidden.
Our choice also eliminates the cubic non-abelian anomaly for the $U(3)$ stack.  The lepton doublet $\ell_L$ is $(1,2)_{1}$, which leaves only the right handed electron to fit. The
$\overline{e}_R$ needs to come from strings stretched between the $U(1)$ stack and itself, giving a field in a $(1,1)_{-2}$ representation (for a $U(N)$ stack, these are in the symmetric representation).  Finally, we can have a scalar Higgs field with the appropriate quantum numbers coming from strings stretching between the $Sp(1)$ and $U(1)$ stacks.  The quiver summarizing the spectrum is in Table 1.

\begin{table}[mqsmpart]
\begin{equation*}
\begin{array}{r|r|r}
{\rm Name}&{\rm Rep.}&Q_Y\\
\hline
q_L&(3,2)_{0}&\frac{1}{6}\\
\hline
\overline{u}_R&({\bar 3},1)_{1}&-\frac{2}{3}\\
\hline
\overline{d}_R&({\bar 3},1)_{-1}&\frac{1}{3}\\
\hline
\ell_L&(1,2)_{1}&-\frac{1}{2}\\
\hline
\overline{e}_R&(1,1)_{-2}&1\\
\hline
\varphi &(1,2)_1&-\frac{1}{2}
\end{array}
\end{equation*}
\caption{\textsf{\protect\small Chiral fermion spectrum of the Minimal Quiver Standard Model, plus Higgs.  Here, $Q_Y\equiv \frac{1}{6} Q_{U(3)}-\frac{1}{2}Q_{U(1)}$}}
\end{table} 

In the end, we have mixed anomalies between the two $U(1)$'s and the $SU(N)$ groups.  It is simple to show that the combination $Q_{U(3)}-3Q_{U(1)}$ has no mixed anomaly, and as expected, the hypercharge $Q_Y\equiv \frac{1}{6}Q_{U(3)}-\frac{1}{2}Q_{U(1)}$ is anomaly free.  However, the $Q_{U(3)}$ (gauged baryon number)  is not anomaly free, and we expect this anomaly to be canceled by a Green-Schwarz mechanism, where an extra coupling to a Ramond-Ramond two-form field or axion to the anomalous $U(1)$ will cancel the anomaly and give a mass to the $U(1)$.  This massive $U(1)$ gauge boson is the only extra particle beyond the standard model which this model predicts, and its mass is dependent on the specific geometry of the string construction. 
 The effective lagrangian for the axionic field participating in the anomaly cancellation mechanism is the following
 \begin{equation}
\frac 12 f_a^2 (\partial_\mu a(x) + g S_\mu)^2 + \sum_i A_i a (x) \frac{\tilde g_i^2}{8\pi^2} \epsilon^{\mu\nu\rho\sigma} \hbox{tr}( F^i_{\mu\nu} F^i_{\rho\sigma})
 \end{equation}
Here $a$ is dimensionless and $A_i$ are coefficients determined by the anomaly, while the vector field $S_\mu$ is the one that has mixed anomalies. In the above equation $S$ has canonical normalization, hence the power of $g$ in the mixing term with the axion.
Under gauge transformation of $S_\mu \to S_\mu -  \partial_\mu \Lambda$, we have that $a$ transforms as $a\to a+ g \Lambda$, so the dimension five axionic coupling to the rest of the gauge degrees of freedom is not gauge invariant.  
 
The anomaly cancels when the one loop effective action induced from integrating out the fermions is added to the tree level dimension five axionic coupling term, and the mixing between $a$ and $S$ is taken into account. There is an equivalent formulation in terms of an antisymmetric
two-index tensor, described in more detail in \cite{GIIQ}.

If one restores units to have $a$ canonically normalized, we find that the mass of $S$ is $m_s\sim g f_a$ and that the dimension five operator related to the anomaly is suppressed by $g^2/f_a\sim g^3/ m_S$. When counting loops, this classical term in the action should be counted as a one loop term (the extra powers of $g$ guarantee this). If tree level renormalizable decays are available, the high dimension operators are suppressed by $g^3$ and their contributions to decay rates can be ignored in a first approximation. At that stage we can also choose the Unitary gauge condition $a=0$. Then we have a standard massive vector field with prescribed couplings to matter. The effective theory will break down at the scale  $f_a/g^2$.

Also notice that since we only have one Higgs doublet,  all of the Yukawa couplings are fully constrained by experimental data.
In our construction all coupling constants of the model, except for the mass
of the extra vector particle, are fixed by the standard model coupling constants. In this sense, 
our minimal model has only one free parameter and is very predictive. These need to be matched at the string scale. Renormalization group running changes the low energy values of coupling constants and adds some small dependence on the UV details of the string model. 
Thus, for the purposes of this paper we will allow three parameters to vary freely to model these effects: the mass of the new gauge boson, the mixing angle between that boson and the original $Z$, and the $\rho$ parameter which describes the deviation of the neutral current lagrangian from the standard model. 
A benchmark model would have the string scale at 10 TeV (the scale at which coupling constants are matched to the string relations) and the mass of the $Z'$ at 1 TeV.  We would then have to account for running of the coupling constants and mixings between 10 TeV and 1 TeV before comparing to data. We will often refer to this benchmark point to make estimates.

\section{The Langragian}
Looking at the quiver and table for the MQSM, we see that the field content is identical to the Standard Model, with the exception of one extra gauge boson corresponding to the extra $U(1)$ in the $U(3)$ gauge group.  In the following, we will call the gauge bosons corresponding to the $SU(3)$ part of the $U(3)$, $C^a_\mu$, where $a$ runs from $1$ to $8$, and $C_\mu$ is the gauge boson for the $U(1)$ part of $U(3)$.  $A^a_\mu$ will likewise be the gauge bosons for $Sp(1)\sim SU(2)$, with $a$ running from $1$ to $3$, and $B_\mu$ will be the gauge boson for the additional $U(1)$ group.  The gauge couplings will be called $g_3,g_2, $ and $g_1$ for the $U(3),Sp(1),$ and $U(1)$ gauge groups respectively.  Also note that the gauge coupling for the $C_\mu$ field (the extra $U(1)$ from the $U(3)$) is actually $g_3/\sqrt{6}$, because of the difference in the canonical normalization of $SU(3)$ and $U(1)$.  As discussed above in the calculation of the mixed anomalies of this theory, there is one combination of the $C_\mu$ and $B_\mu$ fields which is anomaly free that we will associate with the hypercharge.  Accounting for differences in the canonical normalization of $SU(3)$ and $U(1)$, we find that the correct combination is

\begin{align}
Y_\mu&\equiv S_P C_\mu-C_P B_\mu\\
\theta_P&\equiv \tan^{-1}\left(\sqrt{\frac{2}{3}}\frac{g_1}{g_3}\right)
\end{align}
Here, we have abbreviated $S_P\equiv\sin \theta_P, C_P\equiv \cos\theta_P$.  We will continue to use this and $T_P\equiv \tan \theta_P$ for the various angles we will need in the following calculations.
Likewise, there is the orthogonal combination which is anomalous, and has an explicit mass term in the lagrangian from the Green-Schwarz anomaly cancellation mechanism:

\begin{align}
Y'_\mu&\equiv C_PC_\mu+S_PB_\mu
\end{align}

Notice that this process lets us determine the hypercharge coupling constant. We find that
\begin{equation}
\frac 1{g_Y^2}= \frac 1{4 g_1^2}+ \frac 1{6 g_3^2} 
\end{equation}
Since experimentally we have that $g_Y$ is substantially smaller than $g_3$, we find that
$ g_1\sim g_Y/2$. This means that the mixing angle $\theta_P$ is fairly small. Consequently, the field $Y'$ is mostly aligned with $C$ and it couples to quarks with strength comparable to $g_3/\sqrt 6$, while it couples to leptons with a coupling proportional to hypercharge. The coupling to leptons scales like $g_1 \sin\theta_P$, and it is numerically very suppressed. This means that the branching fraction of $Y'$ decaying into leptons is very small compared to the branching fraction into hadrons. This is also the way that $Y'$ couples to the Higgs.

When the theory undergoes  electroweak symmetry breaking, because $Y'$ couples to the Higgs, one gets additional mixing. Thus $Y'$ is not exactly a mass eigenstate.

The eigenstates can be written as

\begin{align}
A_\mu&=S_W A^3_\mu+C_W Y_\mu\\
Z_\mu&=\sigma(C_{Z'}C_WA^3_\mu-C_{Z'}S_WY_\mu+S_{Z'}Y'_\mu)\\
Z'_\mu&=\sigma(-S_{Z}C_WA^3_\mu+S_{Z}S_WY_\mu-C_{Z}Y_\mu')
\end{align}
with
\begin{equation}
\theta_{Z'}\equiv \tan^{-1}\left(\frac{\rho S_W T_P}{\rho-\frac{M_{Z'}^2}{M_Z^2}}\right)
\label{quiver-mixing}
\end{equation} 
\begin{align}
\theta_{Z}&\equiv \tan^{-1}\left(\frac{\rho-1}{\rho S_W T_P}\right)\\
\sigma&=\sec(\theta_Z+\theta_{Z'})
\end{align}
where we've defined $\theta_W\equiv \tan^{-1}\left(\frac{2g_1 c_P}{g_2}\right)$ and the $\rho$ parameter in the standard way, $\rho\equiv \frac{M_W^2}{M_Z^2 C_W^2}$.
Note that $\theta_Z$ and $\theta_{Z'}$ are not independent, as 

\begin{equation}
T_{Z}-\frac{1}{T_{Z'}}=\frac{1}{\rho S_W T_P}\left(\frac{M_{Z'}^2}{M_Z^2}-1\right)
\end{equation} 
Both $\theta_Z$ and $\theta_{Z'}$ go to $0$ as $M_{Z^{\prime}}\to \infty$.  $\sigma$ is chosen so that the $Z$'s have canonically normalized kinetic terms, as will be show below.
We need to invert the eigenstates, and we find that

\begin{align}
A^3_\mu&=S_W A_\mu+C_W(S_{Z'} Z'_\mu+C_{Z}Z_\mu)\\
Y_\mu&=C_W A_\mu-S_W(S_{Z'} Z'_\mu+C_{Z}Z_\mu)\\
Y'_\mu&=-C_{Z'} Z'_\mu-S_{Z}Z_\mu
\end{align}
The kinetic terms for the electro-weak sector in the lagrangian are

\begin{equation}
{\mathcal L}_{\rm kin}=-\frac{1}{4}F^{a\mu\nu}F^a_{\mu\nu}-\frac{1}{4}B^{\mu\nu}B_{\mu\nu}-\frac{1}{4}C^{\mu\nu}C_{\mu\nu}
\end{equation}
where

\begin{align}
F^1_{\mu\nu}&=\partial_\mu A^1_\nu-\partial_\nu A^1_\mu+g_2(A^2_\mu A^3_\nu-A^2_\nu A^3_\mu)\\
F^2_{\mu\nu}&=\partial_\mu A^2_\nu-\partial_\nu A^2_\mu+g_2(A^3_\mu A^1_\nu-A^3_\nu A^1_\mu)\\
F^3_{\mu\nu}&=\partial_\mu A^3_\nu-\partial_\nu A^3_\mu+g_2(A^1_\mu A^2_\nu-A^1_\nu A^2_\mu)\\
B_{\mu\nu}&=\partial_\mu B_\nu-\partial_\nu B_\mu\\
C_{\mu\nu}&=\partial_\mu C_\nu-\partial_\nu C_\mu
\end{align}
Then, expressing this in terms of the physical field basis, after some algebra we find the full kinetic and boson self-interaction langrangian for the gauge bosons is:

\begin{align}
\nonumber {\mathcal L}_{\rm kin}&=-\frac{1}{4}F^{\mu\nu}F_{\mu\nu}-\frac{1}{4}Z^{\mu\nu}Z_{\mu\nu}-\frac{1}{4}Z'{}^{\mu\nu}Z'_{\mu\nu}-D^\dagger{}^\mu W^-{}^\nu D_\mu W^+_\nu+D^\dagger{}^\mu W^-_\nu D_\nu W^+_\mu\\
\nonumber &+ie(F^{\mu\nu}+\cot\theta_W\cos\theta_{Z}Z^{\mu\nu}+\cot\theta_W\sin\theta_{Z'} Z'{}^{\mu\nu})W^+_\mu W^-_\nu\\
\nonumber &-\frac{1}{2}\frac{e^2}{\sin^2\theta_W}(W^+{}^\mu W_\mu^-W^+{}^\nu W^-_\nu -W^+{}^\mu W^+_\mu W^-{}^\nu W^-_\nu)\\
&-\frac{1}{2}\sin(\theta_Z+\theta_{Z'})Z^{\mu\nu}Z'_{\mu\nu} \label{self-gauge}
\end{align}
where 

\begin{equation}
D_\mu=\partial_\mu-ie(A_\mu+\cot\theta_W\cos\theta_{Z} Z_\mu+\cot\theta_W\sin\theta_{Z'}Z'_\mu)
\end{equation}
and we have defined $F_{\mu\nu}=\partial_\mu A_\nu-\partial_\nu A_\mu, Z_{\mu\nu}=\partial_\mu Z_\nu-\partial_\nu Z_\mu, Z'_{\mu\nu}=\partial_\mu Z'_\nu-\partial_\nu Z'_\mu$.  This is almost identical to the standard model, except with an additional $Z'$ term in the covariant derivative and $WWZ'$ interaction.  

To calculate the weak neutral current interaction lagrangian, we need to find

\begin{equation}
g_2A^3_\mu T^3_2+g_1B_\mu Q_{U(1)}+\frac{g_3}{\sqrt{6}}C_\mu Q_{U(3)}.
\end{equation}
Using the definitions of $A_\mu, Z_\mu,$ and $Z'_\mu$ found above, we find that

\begin{align}
\nonumber &=eQA_\mu+\frac{e}{S_WC_W}\left((C_Z-S_WS_ZT_P)T^3_2+(S_WS_ZT_P-C_ZS_W^2)Q-\frac{S_WS_Z}{6S_PC_P}Q_{U(3)}\right)Z_\mu\\
&\quad\quad+\frac{e}{S_WC_W}\left((S_{Z'}-S_WC_{Z'}T_P)T^3_2+(S_WC_{Z'}T_P-S_{Z'}S_W^2)Q-\frac{S_WC_{Z'}}{6S_PC_P}Q_{U(3)}\right)Z'_\mu
\end{align}
where we've defined $e\equiv g_2\sin \theta_W$ as usual, and we take $Q\equiv T^3+Y$ as the electric charge, with $Y\equiv \frac{1}{6}Q_{U(3)}-\frac{1}{2}Q_{U(1)}$.
Note that as the mass of the $Z'$ goes to $\infty$, $\theta_Z$ and $\theta_{Z'}$ go to $0$.  It's clear that in this limit the first two terms reduce to the normal ones for the standard model.  Also note that only the quarks have $Q_{U(3)}\neq 0$. 

Taking this form for the covariant derivative, and extracting charges from the quiver, we can write down the full neutral current weak interaction lagrangian:

\begin{align}
\nonumber {\mathcal L}^{NC}&=\frac{eZ_\mu}{S_WC_W}\left((C_Z-S_WS_ZT_P)J_3^\mu+(S_WS_ZT_P-C_ZS_W^2)J_{\rm EM}^\mu-\frac{S_WS_Z}{6S_PC_P}J^\mu_{\rm NEW}\right)+\\
&\quad \frac{eZ'_\mu}{S_WC_W}\left((S_{Z'}-S_WC_{Z'}T_P)J_3^\mu+(S_WC_{Z'}T_P-S_{Z'}S_W^2)J_{\rm EM}^\mu-\frac{S_WC_{Z'}}{6S_PC_P}J^\mu_{\rm NEW}\right)\\
J^\mu_3&=\frac{1}{2}\bar u_L\gamma^\mu u_L-\frac{1}{2}\bar d_L\gamma^\mu d_L+\frac{1}{2}\bar \nu_L\gamma^\mu\nu_L-\frac{1}{2}\bar e_L\gamma^\mu e_L\\
J^\mu_{\rm EM}&=\frac{2}{3}\bar u\gamma^\mu u-\frac{1}{3}\bar d\gamma^\mu d-\bar e\gamma^\mu e\\
J^\mu_{\rm NEW}&=\bar u\gamma^\mu u+\bar d\gamma^\mu d
\end{align}
This is replicated for each of the $3$ generations. The above neutral lagrangian can be written in a shorter way as

\begin{equation}
\mathcal{L}^{NC}=\frac{e}{4S_{W}C_{W}}\left[ \overline{f}\gamma _{\mu
}\left( g_{v}^{f}-g_{a}^{f}\gamma _{5}\right) fZ^{\mu }+\overline{f}\gamma
_{\mu }\left( g_{v}^{\prime f}-g_{a}^{\prime f}\gamma _{5}\right) fZ^{\mu
\prime }\right] ,  \label{lag-1}
\end{equation}
where the vector and axial couplings associated with $Z_{\mu }$ and $Z_{\mu
}^{\prime }$ are 

\begin{eqnarray}
g_{v,a}^{f} &=&g_{v,a}^{f(SM)}C_{Z}+g_{v,a}^{f(quiv)}S_{Z},  \notag \\
g_{v,a}^{\prime f} &=&g_{v,a}^{f(SM)}S_{Z^{\prime
}}+g_{v,a}^{f(quiv)}C_{Z^{\prime }},  \label{Quiver-couplings}
\end{eqnarray}
with $g_{v,a}^{f(SM)}$ the usual SM neutral couplings and $g_{v,a}^{f(quiv)}$
a new ``Quiver'' term given in tab. \ref{EW-couplings}.
 
Taking into account that $M_{Z^{\prime }}^{2}\gg
M_{Z}^{2},$ it is possible to take $C_{Z,Z^{\prime }}\approx 1,$ so that the
neutral couplings in (\ref{Quiver-couplings}) become

\begin{eqnarray}
g_{v,a}^{f} &\approx &g_{v,a}^{f(SM)}+g_{v,a}^{f(quiv)}S_{Z},  \notag \\
g_{v,a}^{\prime f} &\approx &g_{v,a}^{f(quiv)}+g_{v,a}^{f(SM)}S_{Z^{\prime
}}.  \label{Quiver-couplings-2}
\end{eqnarray}
\begin{table}[tbp]
\begin{center}
\begin{tabular}{|c|c|c|c|c|}
\hline
$Fermion$ & $g_{v}^{f(SM)}$ & $g_{a}^{f(SM)}$ & $g_{v}^{f(quiv)}$ & $%
g_{a}^{f(quiv)}$ \\ \hline\hline
$\nu _{j}$ & $1$ & $1$ & $-T_{P}S_{W}$ & $-T_{P}S_{W}$ \\ \hline
$e_{j}$ & $-1+4S_{W}^{2}$ & $-1$ & $-3T_{P}S_{W}$ & $T_{P}S_{W}$ \\ \hline
$u$ & $1-\frac{8}{3}S_{W}^{2}$ & $1$ & $(T_{P}-\frac{2}{3T_{P}})S_{W}$ & $%
-T_{P}S_{W}$ \\ \hline
$d$ & $-1+\frac{4}{3}S_{W}^{2}$ & $-1$ & $(-T_{P}-\frac{2}{3T_{P}})S_{W}$ & $%
T_{P}S_{W}$ \\ \hline
\end{tabular}%
\end{center}
\caption{\textsf{\protect\small Vector and axial SM couplings and the new
Quiver couplings.}}
\label{EW-couplings}
\end{table}

On the other hand, a small mixing angle between the two neutral currents $%
Z_{\mu }$ and $Z_{\mu }^{\prime }$  could appear with the following mass
eigenstates

\begin{eqnarray}
Z_{1\mu } &=&Z_{\mu }C_{\theta }+Z_{\mu }^{\prime }S_{\theta },  \notag \\
Z_{2\mu } &=&-Z_{\mu }S_{\theta }+Z_{\mu }^{\prime }C_{\theta },
\label{Z-ZPrima-mixing}
\end{eqnarray}
where the mixing angle $\theta $ can be constrained by electroweak
parameters at low energy. Taking a very small mixing angle, we can do C$%
_{\theta }\simeq 1,$ so that the Lagrangian from Eq. (\ref{lag-1}) becomes 

\begin{equation}
\mathcal{L}^{NC}\approx \frac{e}{4S_{W}C_{W}}\left[ \overline{f}\gamma _{\mu
}\left( G_{v}^{f}-G_{a}^{f}\gamma _{5}\right) fZ_{1}^{\mu }+\overline{f}%
\gamma _{\mu }\left( G_{v}^{\prime f}-G_{a}^{\prime f}\gamma _{5}\right)
fZ_{2}^{\mu }\right] ,  \label{lag-2}
\end{equation}%
where the couplings associated with $Z_{1\mu }$ are

\begin{equation}
G_{v,a}^{f}=g_{v,a}^{f(SM)}+\delta g_{v,a}^{f},\qquad \delta
g_{v,a}^{f}=g_{v,a}^{\prime f}S_{\theta }+g_{v,a}^{f(quiv)}S_{Z},
\label{coup2}
\end{equation}%
and the couplings associated with $Z_{2\mu }$ are

\begin{equation}
G_{v,a}^{\prime f}=g_{v,a}^{f(quiv)}-\delta g_{v,a}^{\prime f},\qquad \delta
g_{v,a}^{\prime f}=g_{v,a}^{f}S_{\theta }-g_{v,a}^{f(SM)}S_{Z^{\prime }}.
\label{coup3}
\end{equation}
The mixing angle also produce an additional contribution to the $WWZ^{\prime}$ term in the lagrangian from Eq. (\ref{self-gauge}), which becomes

\begin{equation}
{\mathcal L}_{\rm Z_{2}WW}=ie\cot\theta_W(S_{Z'}-S_{\theta})Z_{2}^{\mu\nu}W^+_\mu W^-_\nu . \label{Z2WW}
\end{equation}

\section{Indirect detection at CERN LEP}

The couplings of the $Z_{1\mu }$ bosons in Eq. (\ref{lag-2}) have the same
form as the SM-neutral couplings, where the vector and axial vector
couplings $g_{v,a}^{SM}$ are replaced by $G_{v,a}=g_{v,a}^{SM}I+\delta
g_{v,a},$ and the couplings $\delta g_{v,a}$ (given by eq. (\ref{coup2}))
contain corrections due to the small $Z_{\mu }-Z_{\mu }^{\prime }$ mixing
angle $\theta $ and the deviation of the $\rho $ parameter. For
this reason all the analytical parameters at the Z-pole have the same SM
form but with small correction factors. In the SM, the partial decay widths
of $Z_{1}$ into fermions $f\overline{f}$ is described by \cite{Yao:2006px,Bernabeu:1991pc}:

\begin{equation}
\Gamma _{f}^{SM}=\frac{N_{c}^{f}G_{f}M_{Z_{1}}^{3}}{6\sqrt{2}\pi }\rho _{f}%
\sqrt{1-\mu _{f}^{2}}\left[ \left( 1+\frac{\mu _{f}^{2}}{2}\right) \left(
g_{v}^{f}\right) ^{2}+\left( 1-\mu _{f}^{2}\right) \left( g_{a}^{f}\right)
^{2}\right] R_{QED}R_{QCD},  \label{partial-decay}
\end{equation}

\noindent where $N_{c}^{f}=1$, 3 for leptons and quarks, respectively. $%
R_{QED}=1+\delta _{QED}^{f}$ and $R_{QCD}=1+(1/2)\left( N_{c}^{f}-1\right)
\delta _{QCD}^{f}$ are QED and QCD corrections given by Eq. (\ref{QCD}) in
appendix \ref{appendixA}, and $\mu _{f}^{2}=4m_{f}^{2}/M_{Z}^{2}$ considers
kinematical corrections only important for the $b$-quark. Universal
electroweak corrections sensitive to the top-quark mass are taken into
account in $\rho _{f}=1+\rho _{t}$ and in $g_{v}^{SM}$ which is written in
terms of an effective Weinberg angle \cite{Yao:2006px}

\begin{equation}
\overline{S_{W}}^{2}=\left( 1+\frac{\rho _{t}}{T_{W}^{2}}\right) S_{W}^{2},
\label{effective-angle}
\end{equation}

\noindent with $\rho _{t}=3G_{f}m_{t}^{2}/8\sqrt{2}\pi ^{2}$. Nonuniversal
vertex corrections are also taken into account in the $Z_{1}\overline{b}b$
vertex with additional one-loop leading terms which leads to $\rho _{b}=1-%
\frac{1}{3}\rho _{t}$ and $\overline{S_{W}}^{2}=\left( 1+\rho
_{t}/T_{W}^{2}+2\rho _{t}/3\right) S_{W}^{2}$ .

Table \ref{tab:observables} from appendix \ref{appendixAA} summarizes some
observables at the $Z$ resonance, with their experimental values from CERN
collider (LEP), SLAC Liner Collider (SLC) and data from atomic parity
violation \cite{Yao:2006px}, the SM predictions, and the expressions predicted by the MQSM model.
We use $M_{Z_{1}}=91.1876$ $GeV$, $S_{W}^{2}=0.23113$, and for the predicted
SM partial decay given by (\ref{partial-decay}), we use the values from Eq. (%
\ref{SM-partial-decay}) (see appendix \ref{appendixAA}).

The MQSM predictions from table \ref{tab:observables} in appendix \ref%
{appendixAA} are expressed for the LEP Z-pole observables in terms of SM
values corrected by

\begin{eqnarray}
\delta _{Z} &=&\frac{\Gamma _{u}^{SM}}{\Gamma _{Z}^{SM}}(\delta _{u}+\delta
_{c})+\frac{\Gamma _{d}^{SM}}{\Gamma _{Z}^{SM}}(\delta _{d}+\delta _{s})+%
\frac{\Gamma _{b}^{SM}}{\Gamma _{Z}^{SM}}\delta _{b}+3\frac{\Gamma _{\nu
}^{SM}}{\Gamma _{Z}^{SM}}\delta _{\nu }+3\frac{\Gamma _{e}^{SM}}{\Gamma
_{Z}^{SM}}\delta _{\ell };  \notag \\
\delta _{had} &=&R_{c}^{SM}(\delta _{u}+\delta _{c})+R_{b}^{SM}\delta _{b}+%
\frac{\Gamma _{d}^{SM}}{\Gamma _{had}^{SM}}(\delta _{d}+\delta _{s});  \notag
\\
\delta _{\sigma } &=&\delta _{had}+\delta _{\ell }-2\delta _{Z};  \notag \\
\delta A_{f} &=&\frac{\delta g_{V}^{f}}{g_{V}^{f}}+\frac{\delta g_{A}^{f}}{%
g_{A}^{f}}-\delta _{f},  \label{shift1}
\end{eqnarray}

\noindent where for the light fermions

\begin{equation}
\delta _{f}=\frac{2g_{v}^{f}\delta g_{v}^{f}+2g_{a}^{f}\delta g_{a}^{f}}{%
\left( g_{v}^{f}\right) ^{2}+\left( g_{a}^{f}\right) ^{2}},  \label{shift2}
\end{equation}

\noindent while for the $b$-quark

\begin{equation}
\delta _{b}=\frac{\left( 3-\beta _{K}^{2}\right) g_{v}^{b}\delta
g_{v}^{b}+2\beta _{K}^{2}g_{a}^{b}\delta g_{a}^{b}}{\left( \frac{3-\beta
_{K}^{2}}{2}\right) \left( g_{v}^{b}\right) ^{2}+\beta _{K}^{2}\left(
g_{a}^{b}\right) ^{2}}.  \label{shift3}
\end{equation}

\noindent The above expressions are evaluated in terms of the effective
Weinberg angle from Eq. (\ref{effective-angle}).

The weak charge is written as

\begin{equation}
Q_{W}=Q_{W}^{SM}+\Delta Q_{W}=Q_{W}^{SM}\left( 1+\delta Q_{W}\right) ,
\label{weak}
\end{equation}%
where $\delta Q_{W}=\Delta Q_{W}/Q_{W}^{SM}$. The deviation $\Delta Q_{W}$
which contains new physics is

\begin{eqnarray}
\Delta Q_{W} &=&-16\left[ \left( 2Z+N\right) \left( g_{a}^{e(SM)}\delta
g_{v}^{u}+\delta g_{a}^{e}g_{v}^{u(SM)}\right) +\left( Z+2N\right) \left(
g_{a}^{e(SM)}\delta g_{v}^{d}+\delta g_{a}^{e}g_{v}^{d(SM)}\right) \right]  
\notag \\
&&-16\left[ \left( 2Z+N\right) g_{a}^{\prime e}g_{v}^{\prime u}+\left(
Z+2N\right) g_{a}^{\prime e}g_{v}^{\prime d}\right] \frac{M_{Z}^{2}}{%
M_{Z^{\prime }}^{2}}.  \label{new}
\end{eqnarray}

With the expressions for the Z-pole observables and the experimental data
shown in table \ref{tab:observables}, we perform a $\chi ^{2}$ fit at 95\%
CL, where the free quantities $S_{\theta },$ $M_{Z^{\prime }}$ and $\rho $
can be constrained at the $Z$ peak. We assume a covariance matrix with
elements $V_{ij}=\lambda _{ij}\sigma _{i}\sigma _{j}$ among the Z-pole
observables$,$ with $\lambda $ the correlation matrix and $\sigma $ the quadratic
root of the experimental and SM errors. The $\chi ^{2}$ statistic with three
degrees of freedom (d.o.f) is defined as

\begin{equation}
\chi ^{2}(S_{\theta },M_{Z^{\prime }},\rho )=\left[ \mathbf{y}-\mathbf{F}%
(S_{\theta },M_{Z^{\prime }},\rho )\right] ^{T}V^{-1}\left[ \mathbf{y}-%
\mathbf{F}(S_{\theta },M_{Z^{\prime }},\rho )\right] ,  \label{chi}
\end{equation}

\noindent where $\mathbf{y=\{}y_{i}\mathbf{\}}$ represent the 22
experimental observables from table \ref{tab:observables}, and $\mathbf{F}$
the corresponding MQSM prediction. Table \ref{tab:correlation} from appendix %
\ref{appendixAA} display the symmetrical correlation matrices taken from Ref. \cite{Schael:1995ch}.

At three d.o.f, we get 3-dimensional allowed regions in the ($S_{\theta
},M_{Z_{2}},\rho $) space, which correspond to $\chi ^{2}\leq \chi _{\min
}^{2}+7.815,$ with $\chi _{\min }^{2}=18.93$. First of all, we find the best allowed region in the plane $S_{\theta }-M_{Z^{\prime }}$ for the central value $\rho =1,$ which is displayed in Fig. %
\ref{fig:figure2}. We can see that the lowest bound of $M_{Z^{\prime }}$ decreases when the mixing angle ($S_{\theta}$) increases, while the allowed interval becomes very thin. This behavior is to be expected as we note that the only parameter that imposes a lower bound on the $Z^{\prime}$ mass is the weak charge through Eq. (\ref{new}). We see that the additional contribution $\Delta Q_{W}$ increases when $M_{Z^{\prime}}$ decreases, which reduces the size of the allowed window for new physics in order to keep deviations from the SM. We also see that the plot is not symmetrical with respect to the sign of the mixing angle due to fact that $\Delta Q_{W}$ has a linear dependence on $S_{\theta}$. In this case, we see that negative mixing angles are highly restricted. In general, we find very stringent limits on the mixing angle with values in the region $-2 \times 10^{-4}<S_{\theta}<10^{-3}$ within the range $700$ GeV $<M_{Z^\prime}<10000$ GeV. Fig. \ref{fig:figure3} displays the allowed region in the $\rho
-M_{Z^{\prime }}$ plane for $S_{\theta }=0$. We can see in this case that the allowed region is very sensitive to small variations of the parameter $\rho$. At $\rho =1,$ the lower bound of $M_{Z^{\prime }}$ is 2 TeV; however, this bound decreases abruptly to about 400 GeV when $\rho$ changes to $\rho =1 \pm 0.00018$. This result is within the $2\sigma $ range $0.9993\leq \rho \leq 1.0026$ from electroweak global fits \cite{Yao:2006px}.

\section{Direct detection at Fermilab Tevatron and CERN LHC}

Signals for a $Z^{\prime}$ boson can be searched for in $p\overline{p}$ collisions at Fermilab Tevatron and $pp$ collisions at CERN-LHC. In lepton colliders, indirect limits on $Z^{\prime}$ properties are based on statistical deviations from the SM predictions, and are therefore more sensitive to having the systematic errors under control. This is in contrast to hadronic colliders where $Z^{\prime}$ signals come from direct production, but the sensitivity is dependent on the background associated to the hadronic enviroment. Thus, hadronic reactions are much less sensitive to parameters as small as the $Z-Z^{\prime}$ mixing angle \cite{zprimas}. For example, the decay $Z_{2}\rightarrow W^{+}W^{-}$ can occur through the mixing angle, but this decay mode suffers from the SM background of the associated production of a W and two jets, which could spoil a possible $Z^{\prime}$ signal with very small mixing angles. In particular, from Fig. \ref{fig:figure2} we can see that the MQSM model exhibit a mixing angle of the order $S_{\theta} \sim 10^{-4}$ for $Z^{\prime}$ mass in the TeV region. Thus, for the present analysis, the $Z-Z^{\prime}$ mixing angle can be neglected and
we identify the $Z$ and $Z^{\prime}$ bosons as the physical neutral bosons. On the other hand, we find an additional contribution to the $WWZ_{2}$ coupling through $S_{Z^{\prime}}$, as we can see in Eq. (\ref{Z2WW}). However, from Eq. (\ref{quiver-mixing}) we can estimate $S_{Z^{\prime}} \sim -10^{-4}$ for $M_{Z^{\prime}} = 1$ TeV, which produce a suppression factor to the $Z^{\prime}\rightarrow W^{+}W^{-}$ decay rate.  A standard tree level calculation of the $W$ decay mode shows that $\Gamma^{W^+W^-}_{Z'}=\frac{e^2T_P^2}{192\pi C_W^2}M_{Z'}$ which is suppressed a factor of $T_P^2\approx 0.025$ compared to the fermionic decay channels. Therefore we study the $Z^{\prime}\rightarrow f\overline{f}$ mode which exhibit a bigger signal above the SM backgound than the $W$ mode. 

The differential cross section for the process $pp(p\bar{p})\longrightarrow Z^{\prime
}\longrightarrow f\bar{f}$ is given by \cite{zprimas}

\begin{equation}
\frac{d\sigma }{dMdydz}=\frac{K(M)}{48\pi M^{3}}\sum%
\limits_{q}P[B_{q}G_{q}^{+}(1+z^{2})+2C_{q}G_{q}^{-}z],  \label{difcross}
\end{equation}

\noindent where $M=M_{ff}$ is the invariant final state mass, $z=\cos \theta $ is the
scattering angle between the initial quark and the final lepton in the $Z^{\prime}$ rest frame, $%
K(M)\simeq 1.3$ contains leading QED corrections and NLO QCD
corrections, $y=1/2\log [(E+p_{z})/(E-p_{z})]$ is the rapidity, $E$ is the total
energy, $p_{z}$ is the longitudinal momentum, $P=s^{2}/[(s-M_{Z^{\prime
}}^{2})^{2}+M_{Z^{\prime }}^{2}\Gamma _{Z^{\prime }}^{2}],$ $\sqrt{s}$ is the
collider CM energy, and $M_{Z^{\prime }}$ and $\Gamma _{Z^{\prime }}$ are the $%
Z^{\prime }$ mass and total width, respectively. The parameters $B_{q}=[(%
{g^{\prime}}_{v}^{q})^{2}+({g^{\prime}}_{a}^{q})^{2}][({g^{\prime}}%
_{v}^{f})^{2}+({g^{\prime}}_{a}^{f})^{2}]$ and $C_{q}=4({g^{\prime}}%
_{v}^{q}{g^{\prime} }_{a}^{q})({g^{\prime} }_{v}^{f}{g^{\prime} }_{a}^{f})$
contain the couplings from Eq. (\ref{Quiver-couplings-2}) for the initial quarks $q$ and the
final fermions $f$, while the parameter $G_{q}^{\pm
}=x_{A}x_{B}[f_{q/A}(x_{A})f_{\overline{q}/B}(x_{B})\pm f_{q/B}(x_{B})f_{%
\overline{q}/A}(x_{A})]$ contains the Parton Distribution Functions (PDFs) $%
f(x),$ and the momentum fraction $x=Me^{\pm y}/\sqrt{z}.$ We can
consider the Narrow Width Approximation (NWA), where the ratio $\Gamma
_{Z^{\prime }}^{2}/M _{Z^{\prime }}^{2}$ is very small, so that the
contribution to the cross section can be separated into the $Z^{\prime}$
production cross section $\sigma(pp(\bar{p})\rightarrow Z^{\prime})$ and the
fermion branching ratio of the $Z^{\prime}$ boson $Br(Z^{\prime}%
\rightarrow f\bar{f})$

\begin{equation}
\sigma(pp(\bar{p})\rightarrow f\bar{f})=\sigma(pp(\bar{p})\rightarrow
Z^{\prime})Br(Z^{\prime}\rightarrow f\bar{f}),  \label{NWA}
\end{equation}

In this section, we use the following parameters

\begin{eqnarray}
\alpha ^{-1}=128.91,\quad S_{W}^{2}=0.223057,\quad \Gamma _{Z^{\prime
}}=0.15M_{Z^{\prime}},\quad S_{\theta}=0, \quad \rho =1 ,
\label{EW-parameters}
\end{eqnarray}

\noindent where the total width $\Gamma _{Z^{\prime }}\approx 0.15M_{Z^{\prime }}$ is estimated from the
analysis performed in the Ref. \cite{Berenstein:2006pk}. The above values imply that $S_Z=0$ and $S_{Z^{\prime}} \sim 5\times 10^{-4}$ at the $M_{Z^{\prime}} \sim 1$ TeV scale. 

\subsection{$Z^{\prime}_{MQSM}$ at Tevatron}

Indirect $Z-$pole constraints on $Z-Z^{\prime}$ mixing allow bounds of the mass of the $Z^{\prime}$ boson as low as a half of the TeV scale, as we can see in Figs. \ref{fig:figure2} and \ref{fig:figure3}. Thus, in principle a $Z^{\prime}$ boson from the MQSM could possibly be observed at Fermilab-Tevatron. Searches for $Z^{\prime}$ bosons have been perfomed by the CDF and DO collaboration in $e^{+}e^{-}$ \cite{:2007sb}, $\mu ^{+}\mu ^{-}$ \cite{mumu}, $e\mu$ \cite{emu}, $\tau ^{+}\tau ^{-}$ \cite{tau}, and $t\overline{t}$ \cite{top} final states. The detection features (luminosity, triggering, kinematical cuts, etc) depends on the decay channel and are described in the above references for each final state. In particular, the $e^{+}e^{-}$ channel is interesting due to it's good mass resolution and large acceptance. A recent report of a search for electron-positron events in the invariant
mass range $150-950$ GeV collected by the CDF II detector at the Fermilab
Tevatron \cite{:2007sb} has excluded possible $Z^{\prime}$ particles for five different
models: the $Z^{\prime}_{\eta}$, $Z^{\prime}_{\chi}$, $Z^{\prime}_{\psi}$
and $Z^{\prime}_{I}$ bosons from the $E_{6}$ model, and the $Z^{\prime}_{SM}$
from the Sequential Standard Model (SSM). 

We extend the above analysis and find limits for the $Z^{\prime}_{MQSM}$ mass from the MQSM model. The detection feature is described in detail in Ref. \cite{:2007sb} for $e^{+}e^{-}$ events based on an integrated luminosity $L=1.3 fb^{-1}$ of $p\overline{p}$ collisions at C.M. energy $\sqrt{s}=1.96$ TeV, where events are required to have two electrons with tranverse energy $E_{T} \geq 25$ GeV. One electron is required to be incident in the central calorimeter within the pseudorapidity range $\left| \eta \right| \leq 1.1$ and the other electron is allowed to be incident in either the central calorimeter or the plug calorimeter with $1.2 \leq \left| \eta \right| \leq 3.0$. For this study, we use the CalcHep package \cite{calchep} in order to simulate $p\bar{%
p}\rightarrow e^{+}e^{-}$ events with the above kinematical criteria. Using a non-relativistic Breit-Wigner function and the CTEQ6M PDFs \cite{Pumplin:2002vw}, we perform a numerical
calculation with the parameters from Eq. (\ref{EW-parameters}). Fig. \ref{fig:figure4}, shows the $%
95\%$ CL on $\sigma (p\bar{p}\rightarrow Z^{\prime })Br(Z^{\prime
}\rightarrow f\bar{f})$ extracted from Ref. \cite{:2007sb} which does not exhibit any
significant signal above the SM prediction. In the same plot, we show the
corresponding falling prediction for the $Z^{\prime }$ cross section in the
MQSM. For small invariant masses, we see that the MQSM prediction exceeds the $95\%$ CL
limit. A bound is found at $M_{Z^{\prime }}=620$ GeV, where both
curves cross. Thus, the data collected by Tevatron exclude MQSM $Z^{\prime }$ masses below $620$ GeV, which is between 15\% and 30 \% lower than what has been reported for other models \cite{:2007sb}.   

\subsection{$Z^{\prime}_{MQSM}$ at LHC}

The design criteria of ATLAS and CMS at the LHC could reveal a $Z^{\prime }$ signal at the
TeV scale. A collection of signatures for physics beyond the SM is reviewed in Ref. \cite{hp0802.3715} according to the experimental objects that appear in the final state. In particular, the dilepton channels exhibit numerous advantages including easy triggering, low instrumental and SM backgrounds, and well known (NNLO) SM cross sections. Thus, we consider at first this channel to probe $Z^{\prime}$ bosons from MQSM. For leptons, the $\ell =e,\mu $ channel, we use the kinematical cuts reported in Ref. \cite{Dittmar:2003ir} and \cite{hp0801.4389}, based on an integrated luminosity $L=100 fb^{-1}$ of $pp$ collisions at C.M. energy $\sqrt{s}=14$ TeV, where events are required to have two leptons with tranverse energy $E_{T} \geq 20$ GeV within the pseudorapidity range $\left| \eta \right| \leq 2.5$. Additionally, the lepton should be isolated within a cone of angular radius $\Delta R = 0.5$ around the lepton \cite{Dittmar:2003ir}. We use the same parameters from Eq. (\ref{EW-parameters}). Fig. \ref{fig:figure5}a shows the invariant mass distribution for the di-electron system as
final state, where we have chosen a central value $M_{Z^{\prime }}=1000$ GeV. We also show the SM Drell-Yan spectrum in the same plot. We can see that the $Z^{\prime}$ signal exhibits a slightly wide peak above the SM background with about 10 events/GeV. The total cross section is shown in Fig. \ref{fig:figure5}b as a function of the $Z^{\prime}$ mass in the range $600$ GeV $\leq M_{Z^{\prime}}\leq 5000$ GeV. We can get the number of events using the conversion factor $N=\sigma L$, where $L$ is the luminosity $L=100$ $fb^{-1}$. Thus, for $M_{Z^{\prime}}=1000$ GeV we find about $1500$ events, and this number of events decreases to $1$ event when $M_{Z^{\prime}}\sim 5000$ GeV, which confirms the typical $Z^{\prime}$ boson LHC discovery potential.

Another interesting channel is $Z^{\prime}\rightarrow t\overline{t}$, due to the large number of events that will be produced at LHC (about 80 million pair events). Once the LHC precisely determine the top quark properties, $t\overline{t}$ production will offer an excellent chance to search for new physics in the multi-TeV region. Although the sensitivity to new gauge bosons from quark pairs is reduced compared to lepton pairs due to the QCD background, a number of models have a preferential coupling to top quarks \cite{hp0802.3715}, \cite{hp0405055}, \cite{hp0707.2066}, which enhance the new physics signal. For example, the MQSM model exhibit stronger vector couplings to quarks than leptons as we can verify in Tab. 2 due to the enhancement factor $2/3T_{p}$. Thus, we find a large ratio $g_{v}^{u(quiv)}/g_{v}^{e(quiv)}\sim 15$. We are interested in signals for $Z^{\prime}$ bosons in the $t\overline{t}$ mode as final states using the basic cuts $p_{T}(t)\geq 100$ GeV for transverse momentum within the rapidity range $\left| y(t) \right| \leq 2.5$ \cite{hp0405055}, \cite{hp0707.2066}. A detailed discussion about the high invariant mass $t\overline{t}$ event reconstruction from different top decay modes can be found in Refs. \cite{hp0802.3715} and \cite{hp0707.2066}. Fig. \ref{fig:figure6}a shows the distribution for $pp\rightarrow t\overline{t}$, where a small signal above the background can be identified for $M_{Z^\prime}=1000$ GeV. We also show the $p_{T}(t)$ distribution in Fig. \ref{fig:figure6}b, which peaks at about $M_{Z^{\prime}}/2=500$ GeV. The huge number of quark top production expected at LHC is confirmed in Fig. \ref{fig:figure6}c corresponding to the total cross section, where we observe about $10^{6}$ events for a luminosity of $L=100$ $fb^{-1}$ and $M_{Z^\prime}=1000$ GeV in the framework of the MQSM model.     

Finally, the bottom quark channel $Z^{\prime}\rightarrow b\overline{b}$ exhibit better kinematical conditions than the top channel, due to the large mass relation $m_{t}/m_{b} \sim 40$. However, the relative width for the decays of the $Z^{\prime}$ into $t\overline{t}$ and $b\overline{b}$ becomes near $1$ for large $M_{Z^{\prime}}$ values. The cross section of the signals depend of the branching ratio of $Z^{\prime}$, as we can see in Eq. (\ref{NWA}). In particular, for the top and bottom signals, we find the following relation \cite{hp0405055}

\begin{equation}
\frac{\Gamma (Z^{\prime}\rightarrow t\overline{t})}{\Gamma (Z^{\prime}\rightarrow b\overline{b})}=\left( \frac{M_{Z^{\prime}}^{2}-4m_{t}^{2}}{M_{Z^{\prime}}^{2}-4m_{b}^{2}} \right) ^{1/2} \left[ \frac{(M_{Z^{\prime}}^{2}+2m_{t}^{2}) \left( g_{v}^{t(quiv)} \right) ^{2}+(M_{Z^{\prime}}^{2}-4m_{t}^{2}) \left( g_{a}^{t(quiv)} \right) ^{2}}{(M_{Z^{\prime}}^{2}+2m_{b}^{2}) \left( g_{v}^{b(quiv)} \right) ^{2}+(M_{Z^{\prime}}^{2}-4m_{b}^{2}) \left( g_{a}^{b(quiv)} \right) ^{2}} \right]
\end{equation}

where the couplings $g_{v}^{q(quiv)}$ and $g_{a}^{q(quiv)}$ are given in Tab. \ref{EW-couplings}. Using the values from Eq. (\ref{EW-parameters}), and $T_{P}^{2}=0.025$, we obtain that $\Gamma (Z^{\prime}\rightarrow t\overline{t})/\Gamma (Z^{\prime}\rightarrow b\overline{b}) \sim 0.85$ or $\Gamma (Z^{\prime}\rightarrow b\overline{b}) \sim 1.18 \Gamma (Z^{\prime}\rightarrow t\overline{t})$ for $M_{Z^{\prime}}=1000$ GeV. Thus, the top signal is about $85\%$ of the bottom signal at $1$ TeV, and will increase for higher $Z^{\prime}$ masses.

\section{Conclusions}

In this paper we have performed a phenomenological study of the properties of the $Z'$ boson in the MQSM model.  The extra boson couples strongly to quarks, essentially as baryon number, and weakly to leptons. It is therefore is less constrained than other similar models by precision electroweak data.  Specifically, we have found that direct searches at the Tevatron exclude masses  less than $620$ GeV, while data from LEP allows a parameter space with the $Z'$ mass greater than $700$ GeV.  We also explored the detection prospects for the $Z'$ at the LHC, and found that at a mass of 1 TeV, there would be a slightly wide peak in the cross section with about $1500$ events for di-electron events, at an integrated luminosity of 100 $fb^{-1}$. We also explored the top quark mode where about $10^{6}$ events are expected. 

The exploration of this model is worthwhile because it is the simplest possible low-energy theory that could arise from a brane-world scenario in a string theory.  Many of the properties of the model, like the existence of extra neutral gauge bosons, are generic across large classes of models using D-branes, and the economy of the model makes it possible to do a full phenomenological analysis.  
Although the prospects for this model being exactly what will be observed at the LHC are slim, hopefully the results here will be useful to future model builders in determining the theory that best explains any upcoming experimental results.

\vspace{0.3cm}

This work was supported by Colciencias, ALFA-EC funds through the HELEN programme, and the Japanese Society for the Promotion of Science. D. B. work supported in part by the U.S. Department of
Energy under grant DE-FG02-91ER40618. D. B. would like to thank the Galileo Galilei Institute for Theoretical Physics and the  Simons Workshop in Mathematical Physics for their hospitality as well as the INFN for partial support during the completion of this work.

\section*{Appendix}

\appendix

\section{Radiative Corrections\label{appendixA}}

The $Z$ decay width in Eq. (\ref{partial-decay}) contains QED and QCD
corrections with $R_{QED}=1+\delta _{QED}^{f}$ and $R_{QCD}=1+(1/2)\left(
N_{c}^{f}-1\right) \delta _{QCD}^{f}$, where \cite{Yao:2006px,Bernabeu:1991pc}

\begin{eqnarray}
\delta _{QED}^{f} &=&\frac{3\alpha Q_{f}^{2}}{4\pi };  \notag \\
\delta _{QCD}^{f} &=&\frac{\alpha _{s}}{\pi }+1.405\left( \frac{\alpha _{s}}{%
\pi }\right) ^{2}-12.8\left( \frac{\alpha _{s}}{\pi }\right) ^{3}-\frac{%
\alpha \alpha _{s}Q_{f}^{2}}{4\pi ^{2}}  \label{QCD}
\end{eqnarray}%
with $\alpha $ and $\alpha _{s}$ the fine and strong structure constants,
respectively.

\section{The Z$_{1}$-pole parameters\label{appendixAA}}

The Z$_{1}$-pole parameters with their experimental values from CERN
collider (LEP), SLAC Liner Collider (SLC) and data from atomic parity
violation taken from Ref. \cite{Yao:2006px}, are shown in table \ref%
{tab:observables}, with the SM predictions and the expressions predicted by
MQSM model. The corresponding correlation matrix from Ref. \cite{Schael:1995ch} is
given in table \ref{tab:correlation}. For the quark masses, at Z-pole, we
use the following values

\begin{eqnarray}
m_{u}(M_{Z_{1}}) &=&2.33_{-0.45}^{+0.42}\;\;MeV;\qquad
m_{c}(M_{Z_{1}})=677_{-61}^{+56}\;\;MeV,  \notag \\
m_{t}(M_{Z_{1}}) &=&181\pm 13\;\;GeV;\qquad
m_{d}(M_{Z_{1}})=4.69_{-0.66}^{+0.60}\;\;MeV,  \notag \\
m_{s}(M_{Z_{1}}) &=&93.4_{-13.0}^{+11.8}\;\;MeV;\qquad
m_{b}(M_{Z_{1}})=3.00\pm 0.11\;\;GeV.  \label{quarks-mass}
\end{eqnarray}

\begin{table}[tbp]
\begin{center}
$%
\begin{tabular}{c|c|c|c}
\hline
Quantity & Experimental Values & Standard Model & MQSM Model \\ \hline\hline
$\Gamma _{Z}$ $\left[ GeV\right] $ & 2.4952 $\pm $ 0.0023 & 2.4968 $\pm $
0.0011 & $\Gamma _{Z}^{SM}\left( 1+\delta _{Z}\right) $ \\ \hline
$\Gamma _{had}$ $\left[ GeV\right] $ & 1.7444 $\pm $ 0.0020 & 1.7434 $\pm $
0.0010 & $\Gamma _{had}^{SM}\left( 1+\delta _{had}\right) $ \\ \hline
$\Gamma _{\left( \ell ^{+}\ell ^{-}\right) }$ $MeV$ & 83.984 $\pm $ 0.086 & 
83.996 $\pm $ 0.021 & $\Gamma _{\left( \ell ^{+}\ell ^{-}\right)
}^{SM}\left( 1+\delta _{\ell }\right) $ \\ \hline
$\sigma _{had}$ $\left[ nb\right] $ & 41.541 $\pm $ 0.037 & 41.467 $\pm $
0.009 & $\sigma _{had}^{SM}\left( 1+\delta _{\sigma }\right) $ \\ \hline
$R_{e}$ & 20.804 $\pm $ 0.050 & 20.756 $\pm $ 0.011 & $R_{e}^{SM}\left(
1+\delta _{had}+\delta _{e}\right) $ \\ \hline
$R_{\mu }$ & 20.785 $\pm $ 0.033 & 20.756 $\pm $ 0.011 & $R_{\mu
}^{SM}\left( 1+\delta _{had}+\delta _{\mu }\right) $ \\ \hline
$R_{\tau }$ & 20.764 $\pm $ 0.045 & 20.801 $\pm $ 0.011 & $R_{\tau
}^{SM}\left( 1+\delta _{had}+\delta _{\tau }\right) $ \\ \hline
$R_{b}$ & 0.21629 $\pm $ 0.00066 & 0.21578 $\pm $ 0.00010 & $%
R_{b}^{SM}\left( 1+\delta _{b}-\delta _{had}\right) $ \\ \hline
$R_{c}$ & 0.1721 $\pm $ 0.0030 & 0.17230 $\pm $ 0.00004 & $R_{c}^{SM}\left(
1+\delta _{c}-\delta _{had}\right) $ \\ \hline
$A_{e}$ & 0.15138 $\pm $ 0.00216 & 0.1471 $\pm $ 0.0011 & $A_{e}^{SM}\left(
1+\delta A_{e}\right) $ \\ \hline
$A_{\mu }$ & 0.142 $\pm $ 0.015 & 0.1471 $\pm $ 0.0011 & $A_{\mu
}^{SM}\left( 1+\delta A_{\mu }\right) $ \\ \hline
$A_{\tau }$ & 0.136 $\pm $ 0.015 & 0.1471 $\pm $ 0.0011 & $A_{\tau
}^{SM}\left( 1+\delta A_{\tau }\right) $ \\ \hline
$A_{b}$ & 0.923 $\pm $ 0.020 & 0.9347 $\pm $ 0.0001 & $A_{b}^{SM}\left(
1+\delta A_{b}\right) $ \\ \hline
$A_{c}$ & 0.670 $\pm $ 0.027 & 0.6678 $\pm $ 0.0005 & $A_{c}^{SM}\left(
1+\delta A_{c}\right) $ \\ \hline
$A_{s}$ & 0.895 $\pm $ 0.091 & 0.9356 $\pm $ 0.0001 & $A_{s}^{SM}\left(
1+\delta A_{s}\right) $ \\ \hline
$A_{FB}^{\left( 0,e\right) }$ & 0.0145 $\pm $ 0.0025 & 0.01622 $\pm $ 0.00025
& $A_{FB}^{(0,e)SM}\left( 1+2\delta A_{e}\right) $ \\ \hline
$A_{FB}^{\left( 0,\mu \right) }$ & 0.0169 $\pm $ 0.0013 & 0.01622 $\pm $
0.00025 & $A_{FB}^{(0,\mu )SM}\left( 1+\delta A_{e}+\delta A_{\mu }\right) $
\\ \hline
$A_{FB}^{\left( 0,\tau \right) }$ & 0.0188 $\pm $ 0.0017 & 0.01622 $\pm $
0.00025 & $A_{FB}^{(0,\tau )SM}\left( 1+\delta A_{e}+\delta A_{\tau }\right) 
$ \\ \hline
$A_{FB}^{\left( 0,b\right) }$ & 0.0992 $\pm $ 0.0016 & 0.1031 $\pm $ 0.0008
& $A_{FB}^{(0,b)SM}\left( 1+\delta A_{e}+\delta A_{b}\right) $ \\ \hline
$A_{FB}^{\left( 0,c\right) }$ & 0.0707 $\pm $ 0.0035 & 0.0737 $\pm $ 0.0006
& $A_{FB}^{(0,c)SM}\left( 1+\delta A_{e}+\delta A_{c}\right) $ \\ \hline
$A_{FB}^{\left( 0,s\right) }$ & 0.0976 $\pm $ 0.0114 & 0.1032 $\pm $ 0.0008
& $A_{FB}^{(0,s)SM}\left( 1+\delta A_{e}+\delta A_{s}\right) $ \\ \hline
$Q_{W}(Cs)$ & $-$72.62 $\pm $ 0.46 & $-$73.17 $\pm $ 0.03 & $%
Q_{W}^{SM}\left( 1+\delta Q_{W}\right) $ \\ \hline
\end{tabular}%
\ \ $%
\end{center}
\caption{\textit{The parameters for experimental values, SM predictions and
MQSM corrections. The values are taken from Ref. \cite{Yao:2006px}}}
\label{tab:observables}
\end{table}

\begin{table}[tbp]
\begin{tabular}{ll}
\hline
$\Gamma _{had}$ & $\Gamma _{\ell }$ \\ \hline\hline
1 &  \\ 
.39 & 1 \\ \hline
\end{tabular}%
\par
\begin{tabular}{lll}
\hline
$A_{e}$ & $A_{\mu }$ & $A_{\tau }$ \\ \hline\hline
1 &  &  \\ 
.038 & 1 &  \\ 
.033 & .007 & 1 \\ \hline
\end{tabular}%
\par
\begin{tabular}{llllll}
\hline
$R_{b}$ & $R_{c}$ & $A_{b}$ & $A_{c}$ & $A_{FB}^{(0,b)}$ & $A_{FB}^{(0,c)}$
\\ \hline\hline
1 &  &  &  &  &  \\ 
-.18 & 1 &  &  &  &  \\ 
-.08 & .04 & 1 &  &  &  \\ 
.04 & -.06 & .11 & 1 &  &  \\ 
-.10 & .04 & .06 & .01 & 1 &  \\ 
.07 & -.06 & -.02 & .04 & .15 & 1 \\ \hline
\end{tabular}%
\par
\begin{tabular}{llllllll}
\hline
$\Gamma _{Z}$ & $\sigma _{had}$ & $R_{e}$ & $R_{\mu }$ & $R_{\tau }$ & $%
A_{FB}^{(0,e)}$ & $A_{FB}^{(0,\mu )}$ & $A_{FB}^{(0,\tau )}$ \\ \hline\hline
1 &  &  &  &  &  &  &  \\ 
-.297 & 1 &  &  &  &  &  &  \\ 
-.011 & .105 & 1 &  &  &  &  &  \\ 
.008 & .131 & .069 & 1 &  &  &  &  \\ 
.006 & .092 & .046 & .069 & 1 &  &  &  \\ 
.007 & .001 & -.371 & .001 & .003 & 1 &  &  \\ 
.002 & .003 & .020 & .012 & .001 & -.024 & 1 &  \\ 
.001 & .002 & .013 & -.003 & .009 & -.020 & .046 & 1 \\ \hline
\end{tabular}%
\caption{\textit{The correlation coefficients for the Z-pole observables}}
\label{tab:correlation}
\end{table}

For the partial SM partial decay given by Eq. (\ref{partial-decay}), we use
the following values taken from Ref. \cite{Yao:2006px}

\begin{eqnarray}
\Gamma _{u}^{SM} &=&0.3004\pm 0.0002\text{ }GeV;\quad \Gamma
_{d}^{SM}=0.3832\pm 0.0002\text{ }GeV;  \notag \\
\Gamma _{b}^{SM} &=&0.3758\pm 0.0001\text{ }GeV;\quad \Gamma _{\nu
}^{SM}=0.16729\pm 0.00007\text{ }GeV;  \notag \\
\Gamma _{e}^{SM} &=&0.08403\pm 0.00004\text{ }GeV.  \label{SM-partial-decay}
\end{eqnarray}

\newpage

\begin{figure}[tbph]
\centering \includegraphics[scale=2]{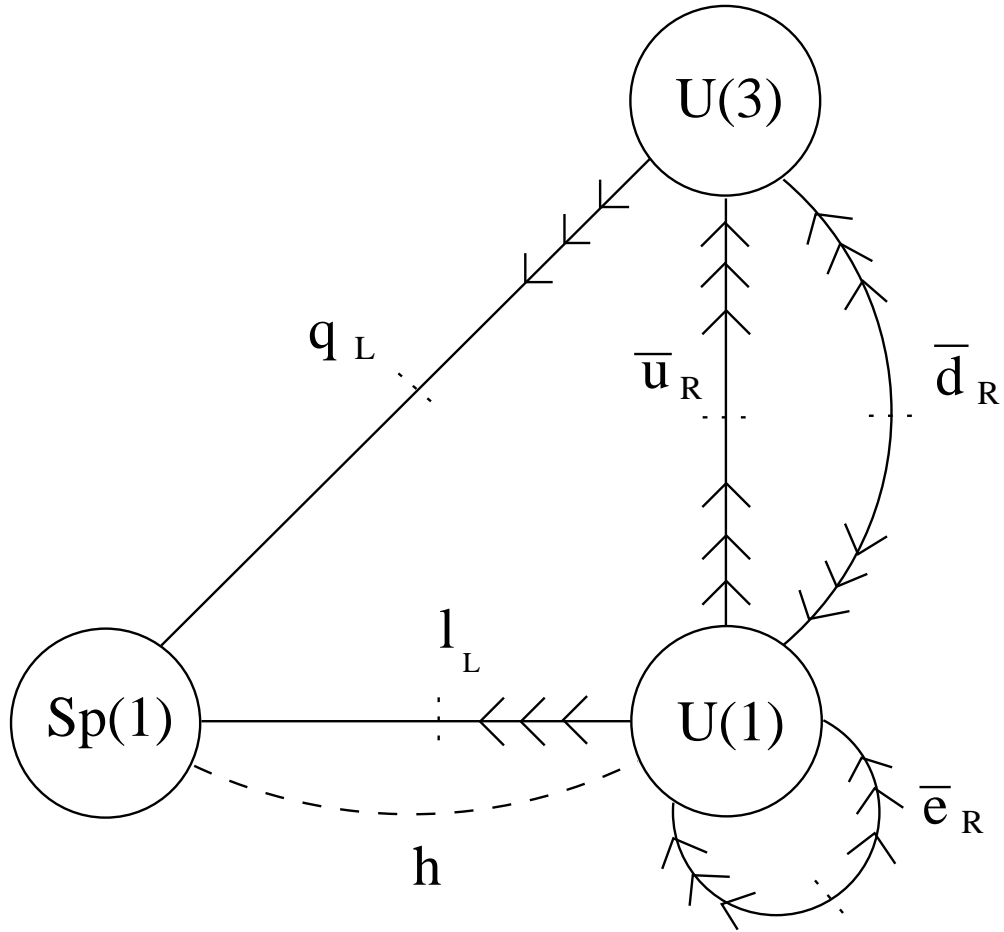}
\caption{\textit{Quiver representation for the Minimal Quiver Standard Model. The arrow directions indicate fundamental or antifundamental representations for the $U(N)$ gauge groups.}}
\label{fig:quiver}
\end{figure}

\begin{figure}[tbph]
\centering \includegraphics[scale=0.8]{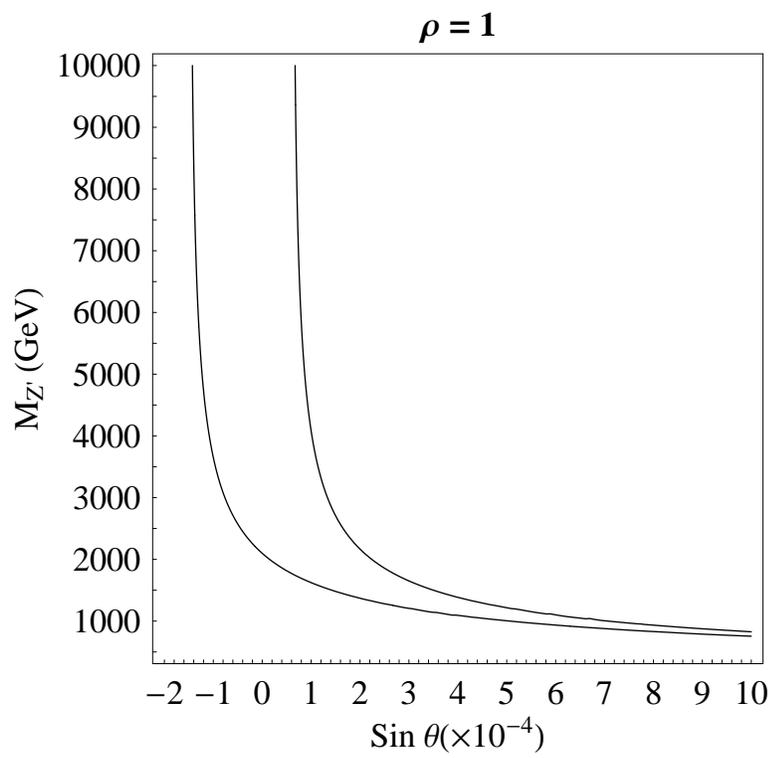}
\caption{\textit{The allowed region for $\sin\protect\theta$ vs $M_{Z^{\prime}}$ with $\rho =1$.}}
\label{fig:figure2}
\end{figure}

\begin{figure}[tbph]
\centering \includegraphics[scale=0.8]{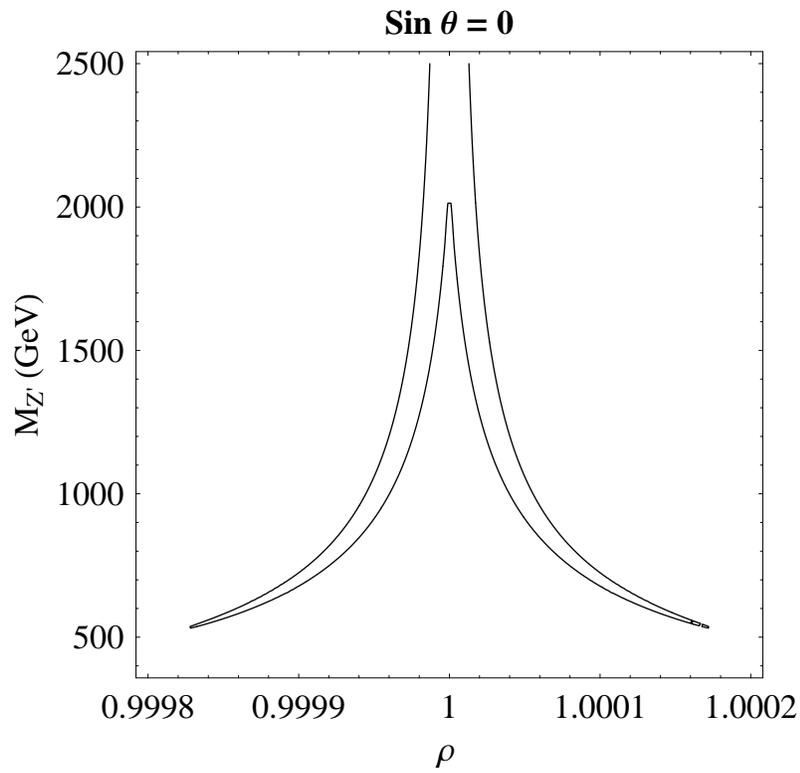}
\caption{\textit{The allowed region for $\rho $ vs $M_{Z^{\prime}}$ with $\sin\protect\theta=0$.}}
\label{fig:figure3}
\end{figure}

\begin{figure}[tbph]
\centering \includegraphics[scale=0.8]{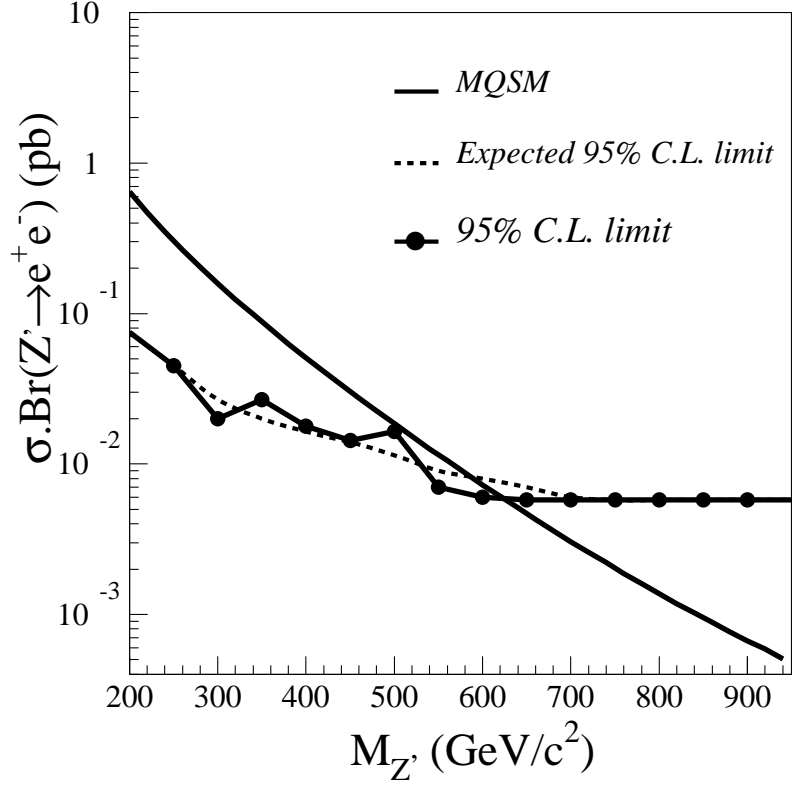}
\caption{\textit{The limit on $\sigma Br$ as a function of the di-electron mass for the $95\%$ CL experimental data in Tevatron and the prediction of the MQSM model. Both plots cross at the bound $620$ GeV.}}
\label{fig:figure4}
\end{figure}

\begin{figure}[tbph]
\centering 
\includegraphics[scale=0.52]{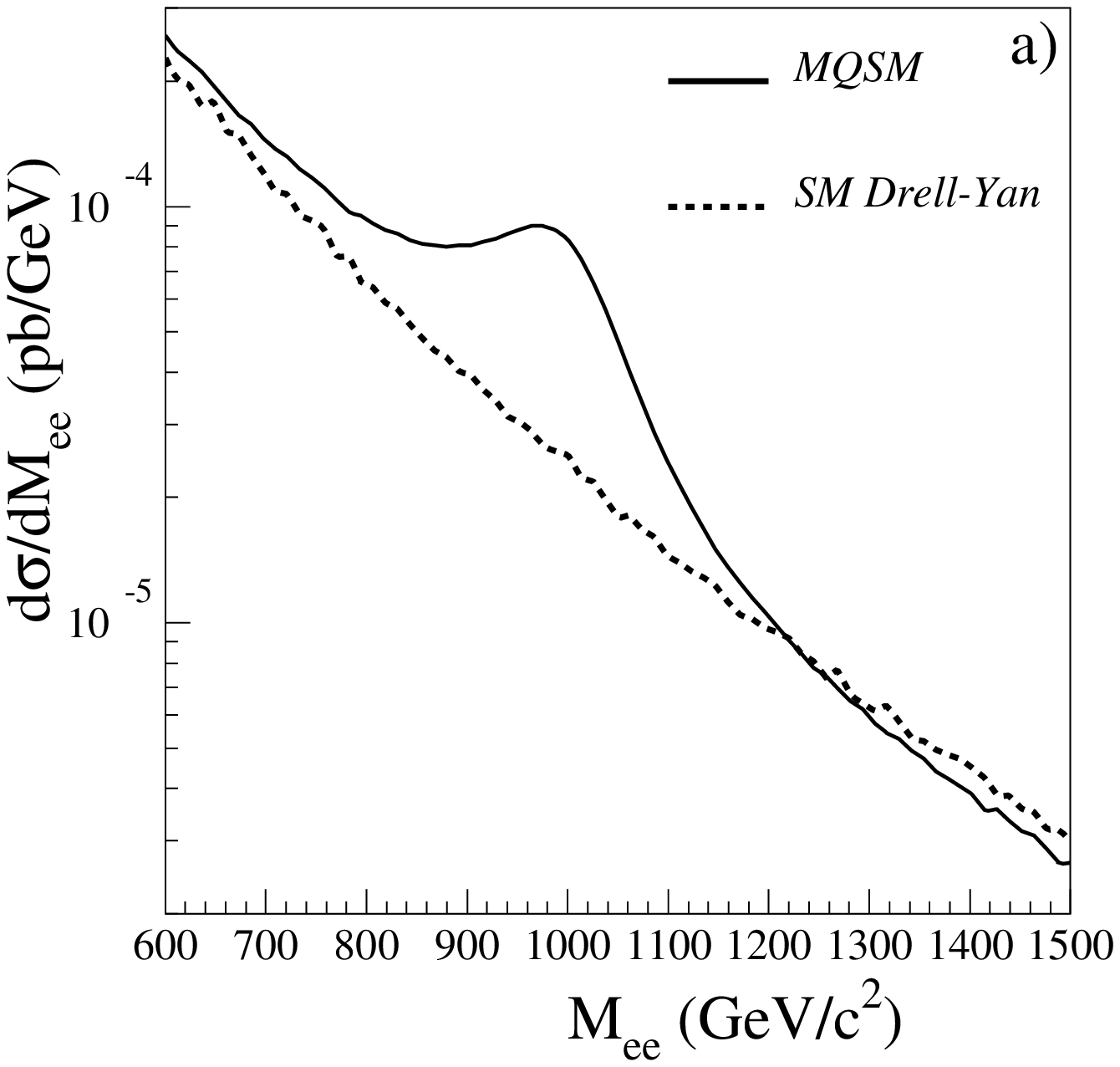} 
\includegraphics[scale=0.52]{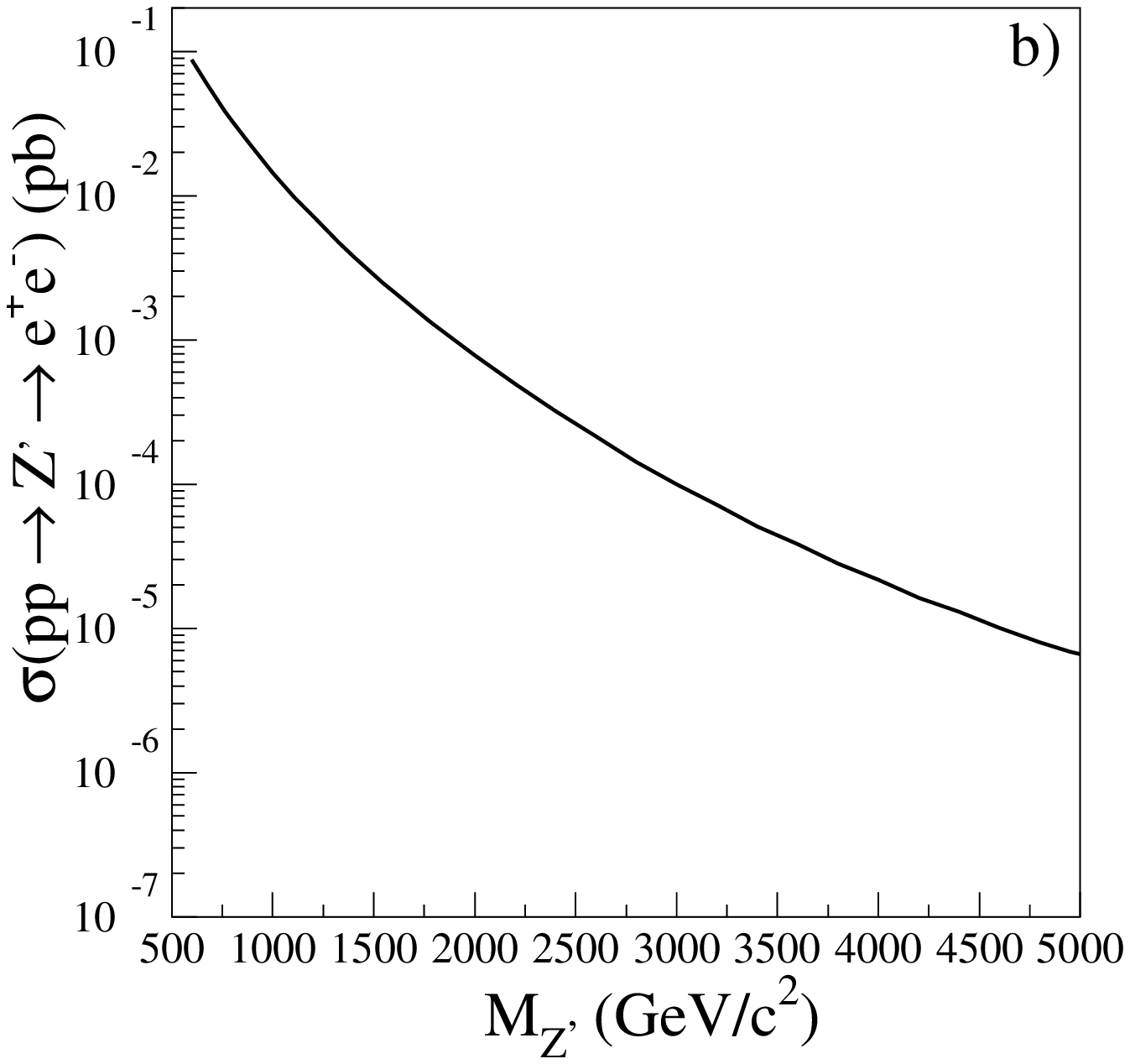}
\caption{\textit{a) Cross section distribution as a function of the electron invariant final state mass for $M_{Z^{\prime}}=1000$ GeV in LHC. b) Total cross section for $e^+e^-$ pair production. The number of events can be obtained using the conversion factor $N=\sigma L$ with $L=100000$ $pb^-1$}}
\label{fig:figure5}
\end{figure}

\begin{figure}[tbph]
\centering 
\includegraphics[scale=0.52]{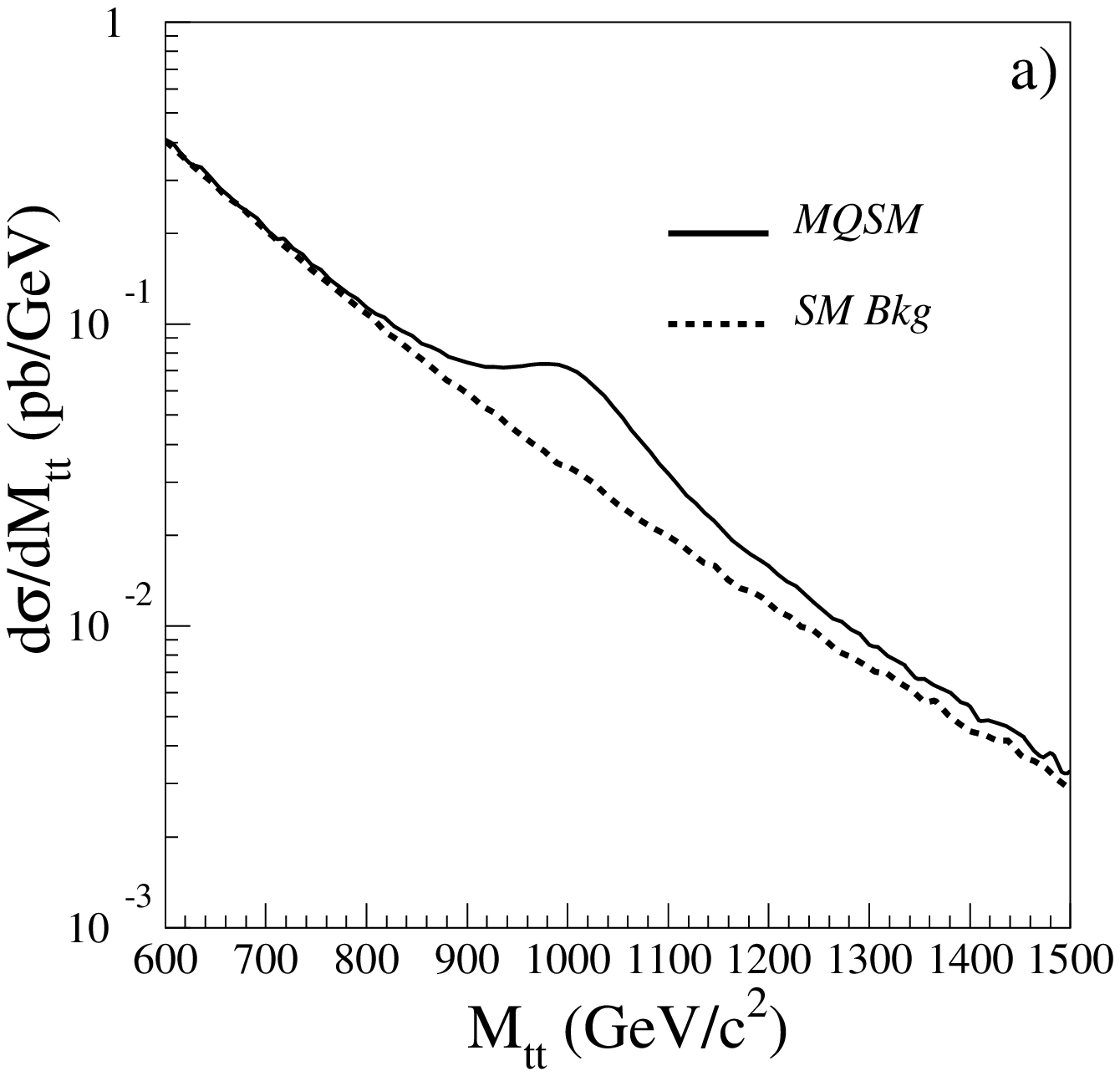} 
\includegraphics[scale=0.52]{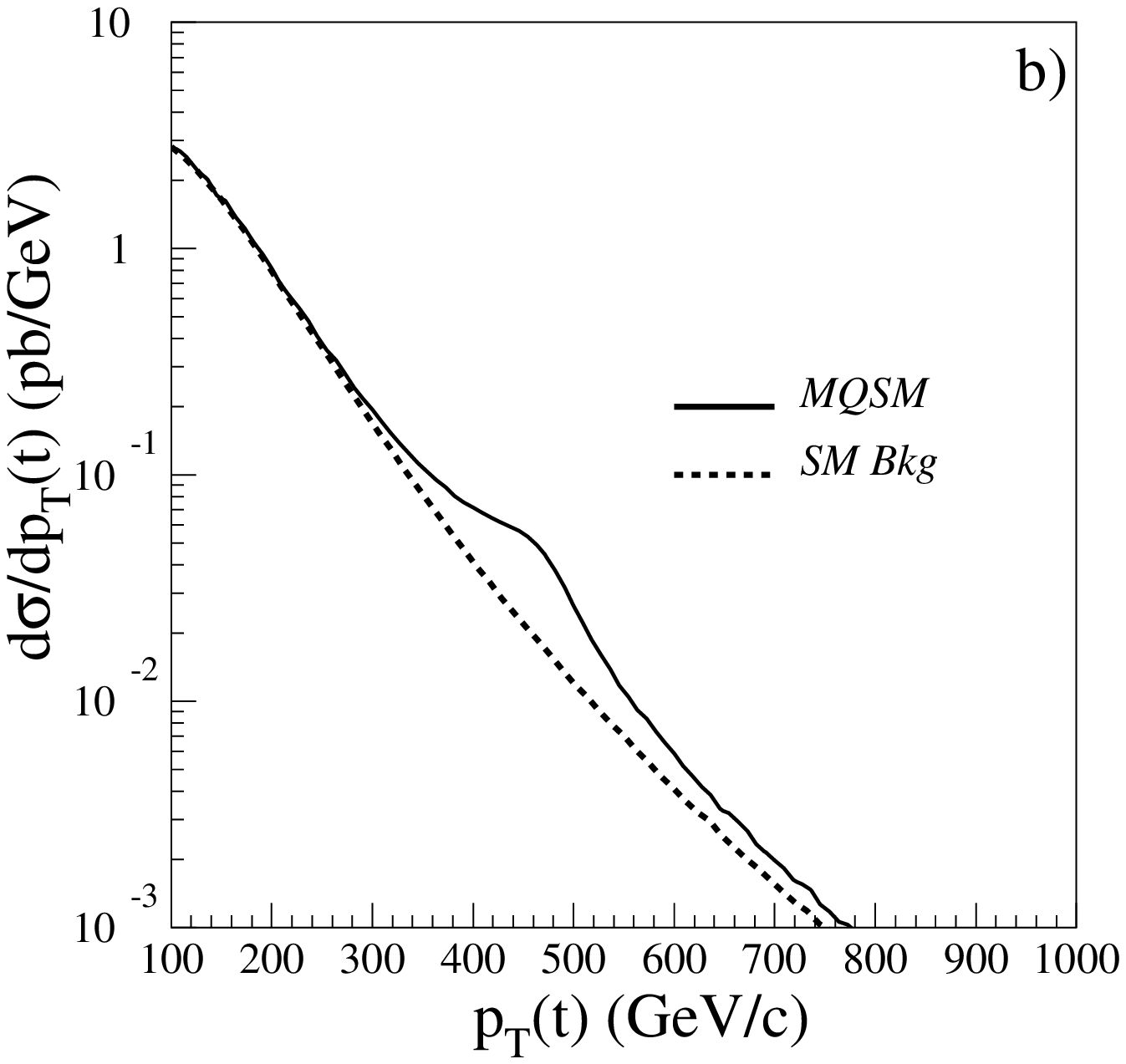}
\includegraphics[scale=0.52]{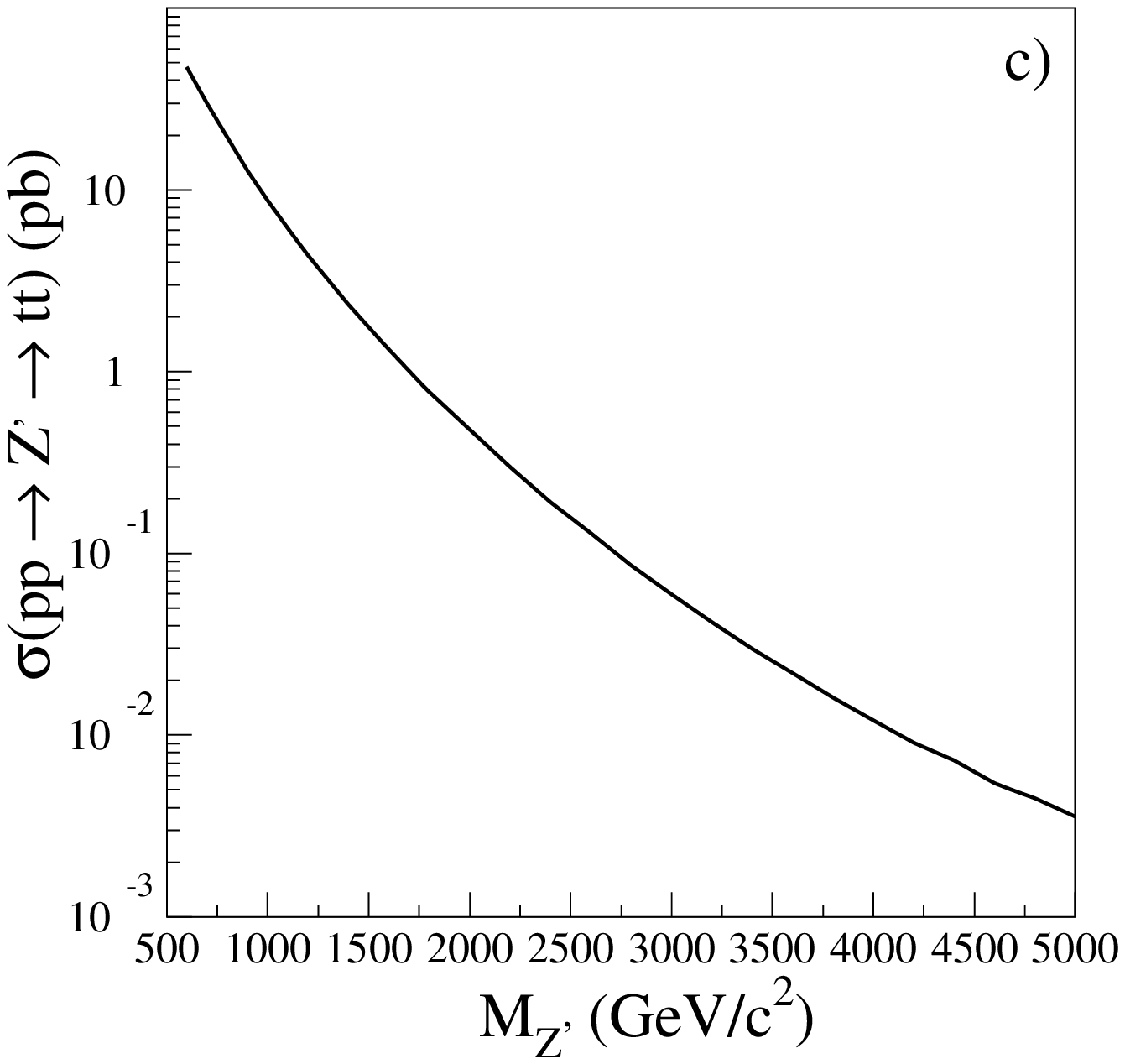}
\caption{\textit{a.)Cross section distribution as a function of the top quark invariant final state mass for $M_{Z^{\prime}}=1000$ GeV in LHC. b.) The top quark transverse momentum distribution. c.) Total cross section for $t\overline{t}$ pair production. The number of events can be obtained using the conversion factor $N=\sigma L$ with $L=100000$ $pb^-1$}}
\label{fig:figure6}
\end{figure}

\end{document}